# Prediction of superconductivity in metallic boron-carbon compounds from 0 to 100 GPa by high-throughput screening

Feng Zheng[1,2], Yang Sun[1*], Renhai Wang[3], Yimei Fang[2], Feng Zhang[4,5], Shunqing Wu[2*], Qiubao Lin[1*], Cai-Zhuang Wang[4,5], Vladimir Antropov[4,5], Kai-Ming Ho[4]

[1]*School of Science, Jimei University, Xiamen 361021, China*
[2]*Department of Physics, OSED, Key Laboratory of Low Dimensional Condensed Matter Physics (Department of Education of Fujian Province), Jiujiang Research Institute, Xiamen University, Xiamen 361005, China*
[3]*School of Physics and Optoelectronic Engineering, Guangdong University of Technology, Guangzhou 510006, China*
[4]*Department of Physics, Iowa State University, Ames, Iowa 50011, United States*
[5]*Ames Laboratory, U.S. Department of Energy, Ames, Iowa 50011, USA*

**Abstract**

Boron carbon compounds have been shown to have feasible superconductivity. In our earlier paper [Zheng *et al*., Phys. Rev. B 107, 014508 (2023)], we identified a new conventional superconductor of $LiB_3C$ at 100 GPa. Here, we aim to extend the investigation of possible superconductivity in this structural framework by replacing Li atoms with 27 different cations under pressures ranging from 0 to 100 GPa. Using the high-throughput screening method of zone-center electron-phonon interaction, we find that ternary compounds like $CaB_3C$, $SrB_3C$, $TiB_3C$, and $VB_3C$ are promising candidates for superconductivity. The consecutive calculations using the full Brillouin zone confirm that they have $T_c$ < 31 K at moderate pressures. Our study demonstrates that fast screening of superconductivity by calculating zone-center electron-phonon coupling strength is an effective strategy for high-throughput identification of new superconductors.

*Email: yangsun@xmu.edu.cn (Y.S.) wsq@xmu.edu.cn (S.W.) lqb@jmu.edu.cn (Q.L.)

# 1. Introduction

The search for novel materials with high-temperature or room-temperature superconductivity is currently one of the most active research fields in condensed-matter physics. Extensive research has been conducted on H-rich compounds to explore possible high $T_c$ superconductors inspired by Ashcroft's suggestion that pressurized hydrogen may become a high-temperature superconductor [1]. Thanks to key advances in crystal structure search and first-principal calculations, several H-rich compounds are predicted to be promising conventional superconductors with high $T_c$ (above 200 K) at high pressure (Megabar pressures) [2-6]. Experimental studies [7-12] later confirmed a presence of superconductivity in some of these systems. Although these remarkable results suggest that the conventional (electron-phonon) mechanism can achieve high $T_c$ superconductivity, the high-pressure conditions required for these H-rich superconductors also limit their practical applications. Therefore, the focus of this field is gradually shifting towards finding new high-$T_c$ materials that can operate at or close to ambient pressure.

Based on the Migdal-Eliashberg phonon-mediated theory, besides H-rich compounds, other light-element compounds such as boron or carbon compounds with strong covalent bonding and low phonon frequencies are also promising candidates for superconductors with high $T_c$ [13-15]. The current record for conventional superconductivity at ambient pressure is boron-doped Q-carbon with 55 K [16]. Another well-known superconductor is $MgB_2$ (39 K), with a layered metal-intercalated structure [13]. Recently, a carbon-boron clathrate $SrB_3C_3$ was successfully synthesized at 57 GPa [17], which can remain stable at atmospheric pressure and was theoretically predicted to be a superconductor with $T_c \sim 40K$ [18-20]. This clathrate $SrB_3C_3$ composed of $sp^3$ bonding represents a new class carbon-boron superconductor with six four-sided and eight six-sided faces ($4^6 6^8$). Using high-throughput density functional theory calculations on 105 $XYB_6C_6$ structures constructed by replacing Sr with two different metals in $SrB_3C_3$, Geng *et al.* proposed 18 new superconductors and $KPbB_6C_6$ was predicted with an ambient-pressure $T_c$ of 88 K [21]. In our previous paper, we also identified a new carbon-boron clathrate LiB$_3$C with eight four-sided faces and four six-sided faces ($4^86^4$) to be a conventional superconductor ($T_c \sim 22$K) at 100 GPa [22]. Considering that element substitution in prototype superconducting structure is a feasible strategy to obtain new superconductors, in this paper, we employed a high-throughput fast screening method of electron-phonon interaction to extend the investigation of possible superconductivity in LiB$_3$C structure framework by replacing Li atoms with 27 different cations at 0, 25, 50, 75 and 100 GPa. This method is highly effective in screening candidates for superconductivity in large materials databases. Ternary compounds like CaB$_3$C, SrB$_3$C, TiB$_3$C, and VB$_3$C are found to be superconductors at moderate pressures.

## 2. Computational methods

The first-principles calculations were performed by using the projector-augmented wave (PAW) [23] representations with density functional theory as implemented in the Vienna *ab initio* simulation package (VASP) [24, 25]. The exchange and correlation energy was treated within the spin-polarized generalized gradient approximation (GGA) and parameterized by Perdew-Burke-Ernzerhof (PBE) formula [26]. Wave functions were expanded in plane waves up to a kinetic energy cutoff of 520 eV. Brillouin-zone integrations were approximated using special *k*-point sampling of the Monkhorst-Pack scheme [27] with a *k*-point mesh resolution of $2\pi \times 0.03 \text{Å}^{-1}$. Lattice vectors and atomic coordinates were fully relaxed until the force on each atom was less than $0.01 \text{ eV} \cdot \text{Å}^{-1}$. The fast screening of electron-phonon coupling (EPC) constant $\lambda_\Gamma$ at the Brillouin zone center was carried out based on the frozen-phonon method [28]. The zone-center phonon was computed by the PHONOPY software [29, 30], with a finer *k*-point sampling grid of $2\pi \times 0.02 \text{Å}^{-1}$ spacing and a criterion of self-consistent calculation 10$^{-8}$ eV. The full Brillouin-zone EPC calculation was performed with the Quantum ESPRESSO (QE) code [31, 32] based on the density-functional perturbation theory (DFPT) [33]. The ultra-soft pseudopotentials from PSLibrary 1.0.0 (high accuracy) [34] for PBE functional were used.

The kinetic energy cutoffs were 75 Ry for wave functions and 576 Ry for potentials. The charge densities were determined on a $k$ mesh of $20 \times 20 \times 20$. The dynamical matrices were calculated on a $q$ mesh of $4 \times 4 \times 4$. The convergence threshold for self-consistency was $1 \times 10^{-12}$ Ry. The Eliashberg spectral function and EPC strength λ in the full Brillouin zone were calculated by using the electron-phonon linewidth. The superconducting $T_c$ is determined by the Allen-Dynes (A-D) equation [35]. The effective screened Coulomb repulsion constant $\mu^*$ was 0.1 for the best-case scenario.

## 3. Results and discussion

### 3.1 Fast evaluation of the possible electron-phonon interactions of MB$_3$C (M = metal element)

The prototype structure LiB$_3$C was predicted to be a superconductor with $T_c$ of 22 K at 100 GPa, which has two PbO-type [36] layers of B$_3$C. Two B$_3$C layers are connected to form B-C cages with eight four-sided faces and four six-sided faces ($4^8 6^4$), as shown in Fig. 1(a). The cages are composed of 18 vertices with alternating C and B atoms, and each cage contains a single Li cation at the center. Then, we substituted the Li site with 27 different metal elements M (M = Li, Na, Mg, Al, K, Ca, Sc, Ti, V, Cr, Mn, Fe, Co, Ni, Cu, Zn, Rb, Sr, Y, Zr, Nb, Mo, Tc, Ru, Rh, Pd, Ag, and Cd). Each structure and LiB$_3$C was relaxed by the structural optimization at 0, 25, 50, 75 and 100 GPa. These optimized structures were further analyzed by the high-throughput screening with the zone-center EPC strength $\lambda_\Gamma$ based on the frozen phonon method [28]. As shown in Fig. 1(b), the dynamically unstable structures are marked with a cross, and dynamically stable phases are colored according to their $\lambda_\Gamma$ (the summation of zone-center EPC of all modes, i.e., $\lambda_\Gamma = \sum_v \lambda_{\Gamma v}$). The reddish coding indicates stronger EPC. We use a sum $\sum_v \lambda_{\Gamma v} \sim 0.3$ threshold to screen out phases with suitable EPC. It can be found that LiB$_3$C has a high EPC at 100 GPa, which has been confirmed to be a superconductor by calculating EPC in the full Brillouin zone in our previous paper [22]. Besides the LiB$_3$C at 100 GPa, we also identify four phases for further study, namely, CaB$_3$C (50 GPa), SrB$_3$C (75 GPa), TiB$_3$C (50, 75 and 100 GPa) and VB$_3$C

(75 and 100 GPa).

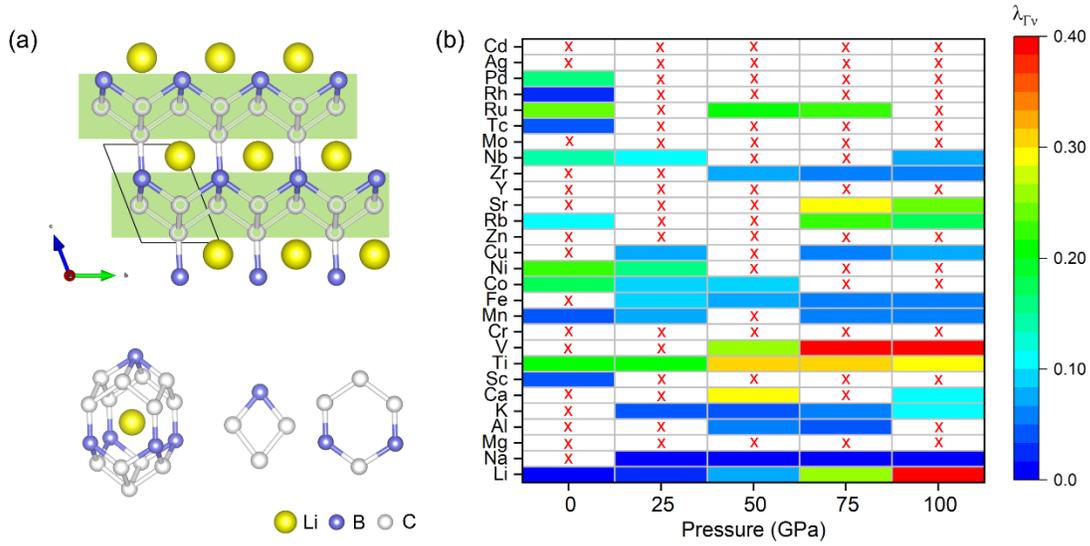

Fig. 1. (a) The prototype structure LiB$_3$C. The light green rectangle denotes PbO-type layers of B$_3$C. (b) The EPC constant $\lambda_\Gamma$ at the Brillouin zone center of MB$_3$C structures at 0, 25, 50, 75, and 100 GPa.

### 3.2 Superconductivity in CaB$_3$C, SrB$_3$C, TiB$_3$C and VB$_3$C

Because of the promising candidate for superconductivity in CaB$_3$C (50 GPa), SrB$_3$C (75 GPa), TiB$_3$C (50, 75, and 100 GPa), and VB$_3$C (75 and 100 GPa), we perform DFPT calculations to compute their full Brillouin zone EPC constant and calculate $T_c$ by the A-D equation.

**CaB$_3$C.** Fig. 2(a) shows the phonon linewidth ($\gamma_{qv}$)-weighted phonon spectrum, projected phonon density of states (PHDOS), and Eliashberg spectral function $\alpha^2F(\omega)$ of CaB$_3$C at 50 GPa. It can be found that the acoustic modes have sizable phonon linewidth, which are mainly attributed to the vibrations of Ca atoms, as indicated by the PHDOS. Furthermore, one can see that phonon modes 4, 5, 9, 12, and 13 also have a large contribution to the phonon linewidth and correspondingly to the electron-phonon interaction at the Γ point, which agrees well with the fast-screening results as shown in Fig. S1(a). Modes 9 and 12 even show large phonon linewidth along the (1/4)Y-Γ-A-(1/3)L$_2$. The vibrational configurations of these modes at the Γ point are shown in Fig. S5. Mode 9 corresponds to stretching vibrations of B2 and B3 atoms and mode 12 undergoes a stretching vibration of C and B1 atoms, which are all along the *c* axis. The integrated EPC parameter λ of CaB$_3$C at 50 GPa is 1.69. The ω$_{log}$ can be obtained from the A-D equation of 237.6 K, and we predict $T_c$ = 31 K (μ$^*$ =0.1). As a comparison, we also compute the EPC for the CaB$_3$C at 100 GPa, as shown in Fig. 2(b), which is not expected to show a strong EPC from zone-center EPC calculation. The integrated EPC parameter λ is 0.73 of CaB$_3$C at 100 GPa. It can be observed that the large phonon linewidth of acoustic modes disappears in CaB$_3$C at 100 GPa. Moreover, only phonon modes 4, 5, 11, and 12 contribute to the phonon linewidth at the Γ point. This also aligns with the fast-screening results, as shown in Fig. S1(b). The vibrational configurations of these modes at the Γ point are shown in Fig. S6, one can see that the vibrations of mode 11 at 100 GPa are identical to those of mode 9 at 50 GPa (Fig. S5). Both correspond to stretching vibrations of B2 and B3 atoms along the *c* axis. This indicates that the original phonon mode 9 undergoes hardening as pressures increase from 50 to 100 GPa of CaB$_3$C. The difference in λ between the CaB$_3$C at 50 and 100 GPa can be explained by their distinct density of states (DOS) at the Fermi level (E$_f$). As shown in Fig. 3(a-b), the electronic properties of CaB$_3$C exhibit metallic features with DOS crossing the E$_f$ at 50 and 100 GPa. The calculated total DOS at the E$_f$ are 1.17 and 0.70 eV$^{-1}$ per f.u. for 50 and 100 GPa, respectively. Such difference in total DOS at the E$_f$ can be attributed to the difference in structural features of CaB$_3$C at 50 and 100 GPa. As shown in Fig. 3 (d), at 50 GPa, two PbO-type layers of B$_3$C (marked with a light green rectangle) in CaB$_3$C are separated by 3.08 Å due to low pressure. The B1 and C1 atoms cannot connect to form a bond, which leads to high DOS at the E$_f$ (Fig. 3 (c)). While, at 100 GPa (Fig. 3 (e)), boron and carbon in CaB$_3$C can retain the prototype structure of LiB$_3$C, which forms a clathrate structure, and two layers of B$_3$C relate to B1 and C1 atoms to form a strong B-C bond (~1.95 Å). This leads to the DOS of B1 and C1 atoms moving towards a lower energy level (Fig. 3 (c)). While due to a larger average phonon frequency (ω$_{log}$ ~ 651.3 K), the $T_c$ of CaB$_3$C at 100 GPa can also reach up to 25 K. The superconducting parameters for CaB$_3$C at 50 and 100 GPa are summarized in Fig. 3(f).

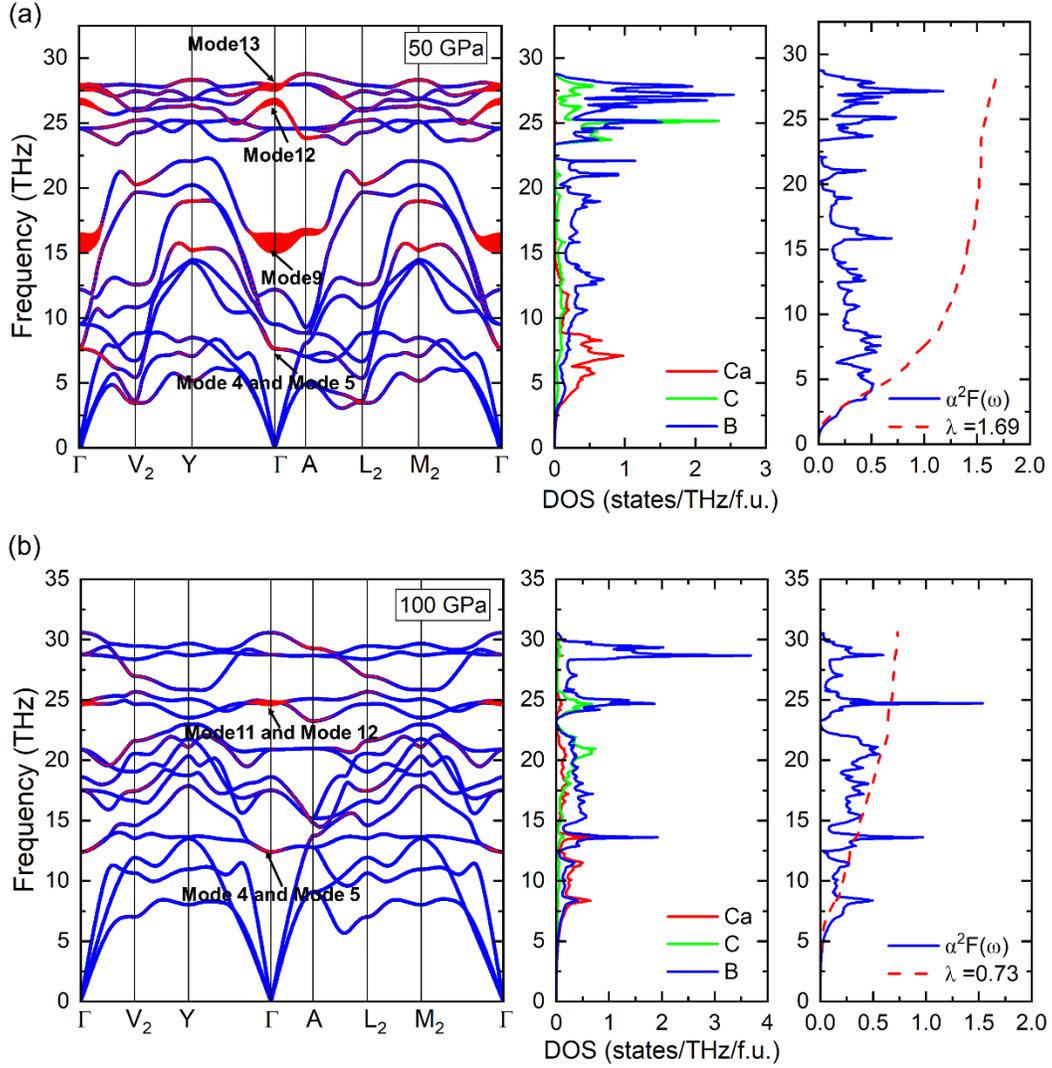

Fig. 2. The $\gamma_{qv}$-weighted phonon spectrum, projected phonon density of states (PHDOS) and Eliashberg spectral function $\alpha^2F(\omega)$ of CaB$_3$C at (a) 50 and (b) 100 GPa.

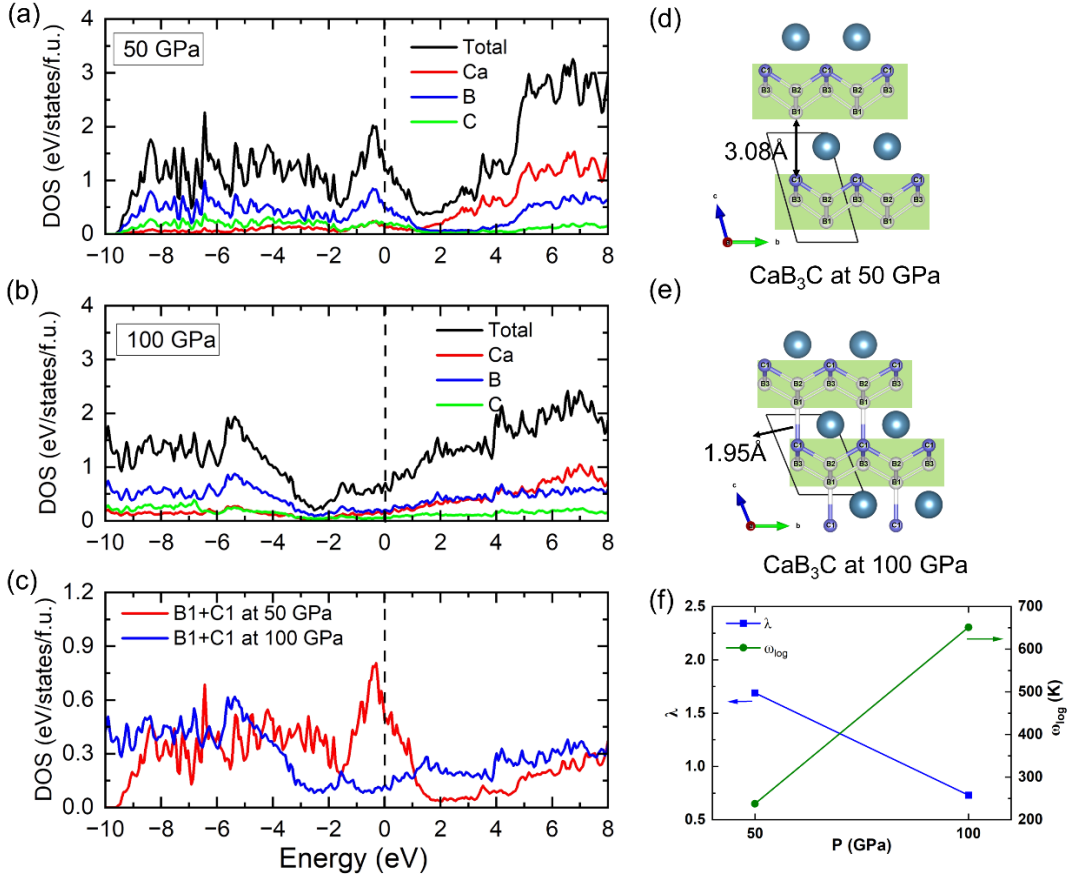

Fig. 3. The electronic density of states (DOS) of CaB$_3$C at (a) 50 and (b)100 GPa. (c) The projected DOS of B1 and C1 atoms at 50 and 100 GPa. Crystal structure of CaB$_3$C at (d) 50 and (e) 100 GPa. The light green rectangle denotes PbO-type layers of B$_3$C. (f) The calculated superconducting parameter for CaB$_3$C at 50 and 100 GPa.

**SrB$_3$C.** Fig.4(a) shows the calculated $\gamma_{qv}$-weighted phonon dispersion and PHDOS of SrB$_3$C at 75 GPa, together with the Eliashberg spectral function α$^2$F(ω). It can be found that the phonon dispersion can be separated into two regions: the lower frequency modes (typically below 7.5 THz) are mainly associated with Sr atoms, and the higher frequency modes can be attributed to boron and carbon atoms. In addition, phonon modes 4, 5, 9, 12, 13, and 14 significantly contribute to the phonon linewidth at the near Γ point. The vibrational configurations of these modes at the Γ point are shown in Fig. S7, which all involve boron and carbon atom vibrations. Like CaB$_3$C at 50 GPa, SrB$_3$C at 75 GPa exhibits large phonon linewidth along the (1/4)Y-Γ-A-(1/3)L$_2$ of phonon modes 9 and 12. However, the values of phonon linewidth are smaller than those of CaB$_3$C at 50 GPa and their frequencies are higher than those of CaB$_3$C at 50 GPa. This leads to a lower total λ of SrB$_3$C at 75 GPa. Our calculated total λ is 1.00, and the ω$_{log}$ is 392.3 K. We estimated $T_c$ = 27 K with μ$^*$ =0.1 by solving the A-D equation. The crystal structure of SrB$_3$C at 75 GPa is shown in Fig. 4(c), like that of CaB$_3$C at 50 GPa. The distance between the two layers of B$_3$C is 3.17 Å. Unlike CaB$_3$C at 100 GPa, SrB$_3$C still has two layers of B$_3$C separated even at 100 GPa (3.07 Å), which can be attributed to the larger ionic radius of Sr. Due to the similar structure, the phonon spectrum of SrB$_3$C at 100 GPa (Fig. 4 (b)) is barely distinguishable from that of 75 GPa (Fig. 4 (a)). It can be found that phonon modes 4, 5, 6, 7, 9, 12, 13, 14, and 15 greatly contribute to the phonon linewidth at the near Γ point. The vibrational configurations of these modes at the Γ point can be seen in Fig. S8, which all involve boron and carbon atom vibrations. The integrated EPC parameter λ of SrB$_3$C at 100 GPa is 0.90, slightly lower than that of 75 GPa. The ω$_{log}$ of SrB$_3$C at 100 GPa is 429.2 K, and $T_c$ is 24 K. The superconducting parameters for SrB$_3$C at 75 and 100 GPa are summarized in Fig. 4(d).

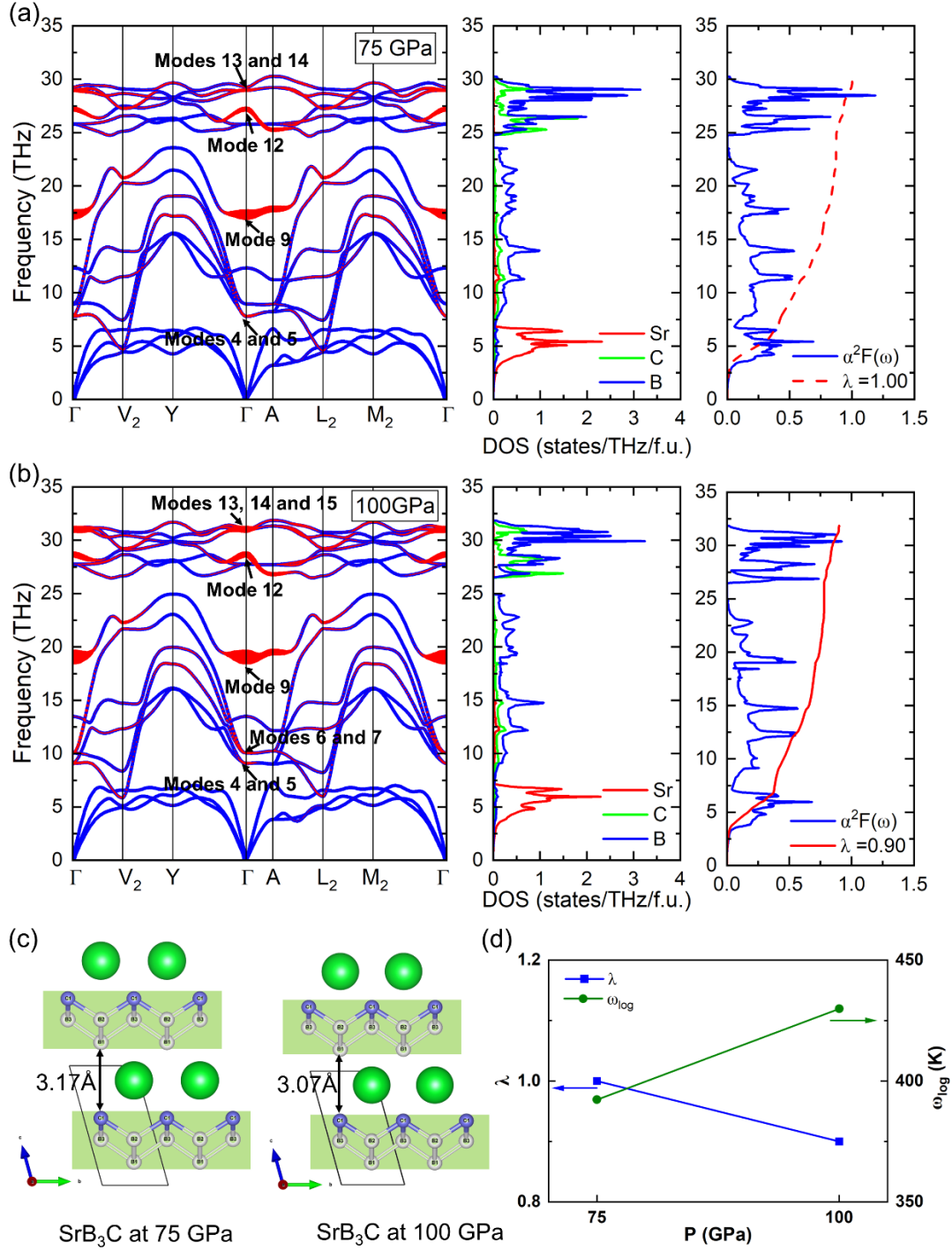

Fig. 4 The $\gamma_{qv}$-weighted phonon spectrum, projected phonon density of states (PHDOS), and Eliashberg spectral function $\alpha^2F(\omega)$ of SrB$_3$C at (a) 75 and (b) 100 GPa. (c) Crystal structure of SrB$_3$C at 75 and 100 GPa. The light green rectangle denotes PbO-type layers of B$_3$C. (d) The calculated superconducting parameter for SrB$_3$C at 75 and 100 GPa.

**TiB₃C and VB₃C.** Fig. 5(a) shows the $\gamma_{qv}$-weighted phonon spectrum, PHDOS, Eliashberg spectral function $\alpha^2F(\omega)$, and crystal structure of TiB$_3$C at 50 GPa. Because of the shift of boron chains, TiB$_3$C forms an irregular B-C structural framework at 50GPa. The EPC is primarily derived from phonon modes at the zone center of TiB$_3$C at 50 GPa. Our calculated integrated EPC parameter λ is 0.80. We obtain the ω$_{log}$ of 507.2 K and predict $T_c$ = 24 K for TiB$_3$C at 50 GPa. The corresponding superconductivity properties of TiB$_3$C at 75 and 100 GPa are shown in Fig. S9. When pressure is up to 75 and 100 GPa, it can be observed that the λ of TiB$_3$C is slightly decreased. The calculated $T_c$ of TiB$_3$C is 23 and 21 K at 75 and 100 GPa, respectively. The structural feature of VB$_3$C is like that of TiB$_3$C, which also cannot retain PbO-type B-C layers, as shown in Fig. 5 (b). The integrated EPC parameter λ of VB$_3$C is 0.97 at 75 GPa, and $T_c$ is 29 K. At 100 GPa, the λ of VB$_3$C is 0.88, as shown in Fig. S10, and we predicted its $T_c$ of 26 K.

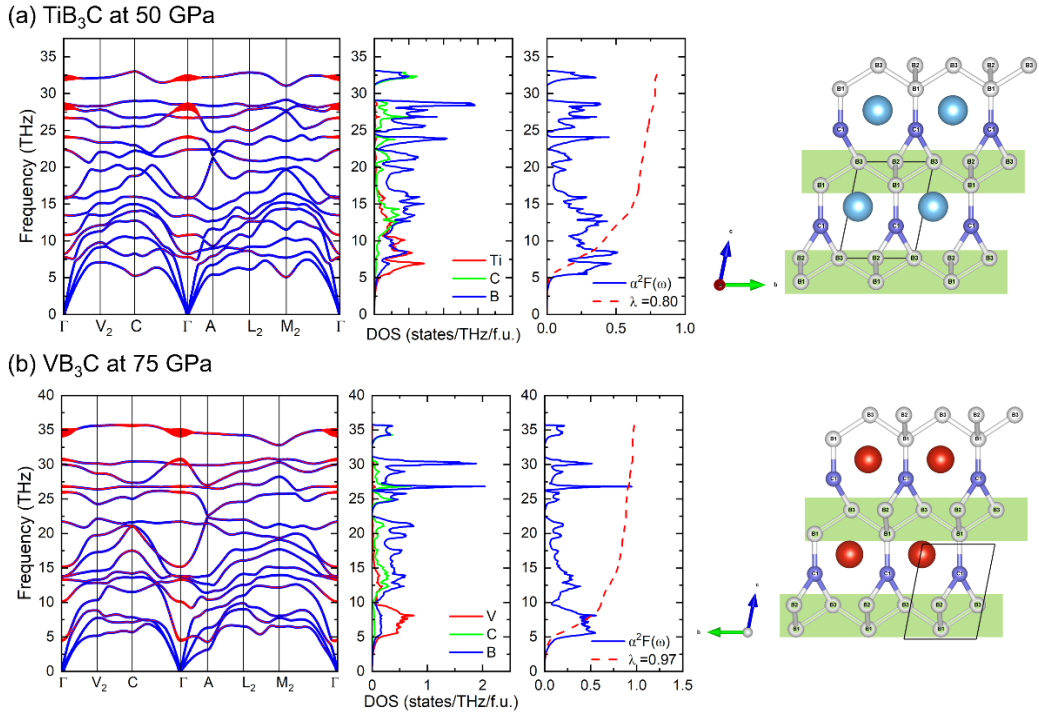

Fig. 5 The $\gamma_{qv}$-weighted phonon spectrum, projected phonon density of states (PHDOS), Eliashberg spectral function $\alpha^2F(\omega)$, and crystal structure of (a) TiB$_3$C at 50 GPa and VB$_3$C at 75 GPa. The light green rectangle denotes the shift of boron chains.

## 4. Conclusions

Using a high-throughput screening method, we performed a broad-range study of the superconductivity of metal-substituted LiB$_3$C. The prototype structure LiB$_3$C was predicted to be a superconductor in our previous paper, which has two PbO-type layers of B$_3$C and two B$_3$C layers connected to form a B-C clathrate structure. Ternary compounds like CaB$_3$C, SrB$_3$C, TiB$_3$C, and VB$_3$C are found to be superconductors at moderate pressures. The CaB$_3$C can retain B-C clathrate structure at 100 GPa and has superconductivity with EPC constant $\lambda$ = 0.73 and $T_c$ of 25 K. While, at 50 GPa, a distance in CaB3C separates two PbO-type B3C layers due to low pressure, which leads to higher DOS at the E$_f$ than that at 100 GPa. Therefore, CaB$_3$C has a higher EPC constant $\lambda$ of 1.69 and $T_c$ = 31 K at 50 GPa. For SrB$_3$C, due to the larger ionic radius of Sr, two PbO-type B$_3$C layers are always separated at 75 and 100 GPa. Calculated results show that SrB$_3$C have $T_c$ of 27 and 24 K at 75 and 100 GPa, respectively. Because of the B chain shift, TiB$_3$C and VB$_3$C cannot retain PbO-type B$_3$C layers and form an irregular B-C structural framework. The EPC constant $\lambda$ is 0.80, and $T_c$ is 24 K for TiB$_3$C at 50 GPa and is slightly decreased with pressure up to 75 and 100 GPa. The VB$_3$C has $\lambda$ of 0.97 and $T_c$ of 29 K at 75 GPa, which also slightly decreases when pressure is up to 100 GPa. The demonstrated consistency between the zone-center electron-phonon interaction calculations and DFPT exploration validates this high-throughput screening approach. This strategy allows a much faster screening process, making it a powerful tool for accelerating the discovery of new functional materials.

## Acknowledgments

F. Zheng was supported by the Research Foundation of Jimei University (ZQ2023013). Work at Xiamen University was supported by the National Natural Science Foundation of China (11874307). Y.S. acknowledges the support of the Fundamental Research Funds for the Central Universities (20720230014). R.W. was supported by the Guangdong Basic and

Applied Basic Research Foundation (Grant No. 2021A1515110328 & 2022A1515012174). V.A., C.Z. Wang, and F. Zhang were supported by the U.S. Department of Energy (DOE), Office of Science, Basic Energy Sciences, Materials Science and Engineering Division. Ames Laboratory is operated for the U.S. DOE by Iowa State University under Contract No. DE-AC02-07CH11358, including the grant of computer time at the National Energy Research Supercomputing Center (NERSC) in Berkeley. S. Fang and T. Wu from the Information and Network Center of Xiamen University are acknowledged for the help with GPU computing.## References

[1] N.W. Ashcroft, Phys. Rev. Lett., **21,** 1748 (1968).
[2] D. Duan, Y. Liu, F. Tian, D. Li, X. Huang, Z. Zhao, H. Yu, B. Liu, W. Tian, T. Cui, Sci. Rep., **4,** 6968 (2014).
[3] F. Peng, Y. Sun, C.J. Pickard, R.J. Needs, Q. Wu, Y. Ma, Phys. Rev. Lett., **119,** 107001 (2017).
[4] H. Liu, Naumov, II, R. Hoffmann, N.W. Ashcroft, R.J. Hemley, Proc. Natl. Acad. Sci., **114,** 6990 (2017).
[5] H. Wang, J.S. Tse, K. Tanaka, T. Iitaka, Y.M. Ma, Proc. Natl. Acad. Sci., **109,** 6463 (2012).
[6] H. Xie, Y. Yao, X. Feng, D. Duan, H. Song, Z. Zhang, S. Jiang, S.A.T. Redfern, V.Z. Kresin, C.J. Pickard, et al., Phys. Rev. Lett., **125,** 217001 (2020).
[7] A.P. Drozdov, M.I. Eremets, I.A. Troyan, V. Ksenofontov, S.I. Shylin, Nature, **525,** 73 (2015).
[8] P. Kong, V.S. Minkov, M.A. Kuzovnikov, A.P. Drozdov, S.P. Besedin, S. Mozaffari, L. Balicas, F.F. Balakirev, V.B. Prakapenka, S. Chariton, et al., Nat. Commun., **12,** 5075 (2021).
[9] E. Snider, N. Dasenbrock-Gammon, R. McBride, X. Wang, N. Meyers, K.V. Lawler, E. Zurek, A. Salamat, R.P. Dias, Phys. Rev. Lett., **126,** 117003 (2021).
[10] M. Somayazulu, M. Ahart, A.K. Mishra, Z.M. Geballe, M. Baldini, Y. Meng, V.V. Struzhkin, R.J. Hemley, Phys. Rev. Lett., **122,** 027001 (2019).
[11] A.P. Drozdov, P.P. Kong, V.S. Minkov, S.P. Besedin, M.A. Kuzovnikov, S. Mozaffari, L. Balicas, F.F. Balakirev, D.E. Graf, V.B. Prakapenka, et al., Nature, **569,** 528 (2019).
[12] L. Ma, K. Wang, Y. Xie, X. Yang, Y. Wang, M. Zhou, H. Liu, X. Yu, Y. Zhao, H. Wang, et al., Phys. Rev. Lett., **128,** 167001 (2022).
[13] J. Nagamatsu, N. Nakagawa, T. Muranaka, Y. Zenitani, J. Akimitsu, Nature, **410,** 63 (2001).
[14] J.M. An, W.E. Pickett, Phys. Rev. Lett., **86,** 4366 (2001).13